\documentclass[conference]{IEEEtran}

\usepackage[english]{babel}
\usepackage[utf8]{inputenc}

\usepackage{amsmath}           

\usepackage{amsthm}
\usepackage{amsfonts}
\usepackage{url}               

\usepackage[pdftex]{graphicx}
\usepackage{epstopdf}
\usepackage{pdfpages}
\usepackage{float}

\usepackage{indentfirst}
\usepackage{algorithmic}
\usepackage{algorithm}
\usepackage{cite}
\usepackage{setspace}

\usepackage{lipsum}

\begin{document}
\title{Massive MIMO Full-Duplex Relaying with Optimal Power Allocation for Independent Multipairs}

\author{
    \IEEEauthorblockN{João S. Lemos\IEEEauthorrefmark{1}, Francisco Rosário\IEEEauthorrefmark{1}, Francisco A. Monteiro\IEEEauthorrefmark{2}, João Xavier\IEEEauthorrefmark{3}, António Rodrigues\IEEEauthorrefmark{1}}
    \IEEEauthorblockA{\IEEEauthorrefmark{1}Instituto de Telecomunicações, and Instituto Superior Técnico, Universidade de Lisboa, Portugal}
    \IEEEauthorblockA{\IEEEauthorrefmark{2}Instituto de Telecomunicações, and ISCTE - Instituto Universitário de Lisboa, Portugal}
    \IEEEauthorblockA{\IEEEauthorrefmark{3}Instituto de Sistemas e Robótica, and Instituto Superior Técnico, Universidade de Lisboa, Portugal}
\{joao.sande.lemos, francisco.rosario\}@tecnico.ulisboa.pt , francisco.monteiro@lx.it.pt
}
\maketitle

\begin{abstract}
With the help of an in-band full-duplex relay station, it is possible to simultaneously transmit and receive signals from multiple users. The performance of such system can be greatly increased when the relay station is equipped with a large number of antennas on both transmitter and receiver sides. In this paper, we exploit the use of massive arrays to effectively suppress the loopback interference (LI) of a decode-and-forward relay (DF) and evaluate the performance of the end-to-end (e2e) transmission. This paper assumes imperfect channel state information is available at the relay and designs a minimum mean-square error (MMSE) filter to mitigate the interference. Subsequently, we adopt zero-forcing (ZF) filters for both detection and beamforming. The performance of such system is evaluated in terms of bit error rate (BER) at both relay and destinations, and an optimal choice for the transmission power at the relay is shown. We then propose a complexity efficient optimal power allocation (OPA) algorithm  that, using the channel statistics, computes the minimum power that satisfies the rate constraints of each pair. The results obtained via simulation show that when both MMSE filtering and OPA method are used, better values for the energy efficiency are attained. 


\end{abstract}

\section{Introduction}
\label{sec:intro}

\looseness -1 In-band full-duplex systems in wireless communications have attracted a lot of research interest in the past few years. This feature is expected to be implemented in future generation networks and gives the possibility of transmitting and receiving data at the same time instant and frequency band, which means that with the same amount of energy as in the half-duplex (HD) counterpart, it is possible to double spectral efficiency \cite{sis}. However, the inherent complexity of such systems has been hindering the 
deployment of full-duplex enabled equipment. The main obstacle to overcome is the difficulty in canceling the signal leakage from the relay output to its input \cite{hien-multi}, \cite{lemos:rls}. Moreover, hardware constraints, algorithms' inefficiency and difficulty in modeling the channel have also posed problems, hence the need for further analysis.

\looseness -1 Alongside the developments in full-duplex, a lot of research has been conducted recently on the capabilities of multiple-input multiple-output (MIMO) systems and its extension to very large arrays (dubbed as massive MIMO). The idea in future generation systems is to use hundreds of antennas at the base station in the hope of serving more users and devices. This study involves various signal processing 
techniques, such as channel modeling and estimation, precoding and detection algorithms, which must be efficient in terms of complexity and perform as close to optimal as possible. 


\looseness -1 In this work, we consider a massive MIMO full-duplex relay architecture, which uses both of the aforementioned state-of-the-art technologies.  We assume a group of $K$ sources communicating with a group of $K$ destinations. This transmission is supported by a decode-and-forward (DF) relay station.

In the first part of this paper, we are interested in loopback interference (LI) mitigation techniques for the considered setup. We assume that both source to relay and LI channels are known (apart from an estimation error) to design a minimum-mean square error filter. This is done in \cite{riihonen2011mitigation} for a small number of antennas. We show that by exploiting the massive MIMO effects and using simple linear processing techniques, LI can be effectively attenuated. The results are evaluated in terms of end-to-end (e2e) bit error rate (BER) performance. Furthermore, an algorithm to find the optimal power allocation (OPA) that fulfils the needs of each communication pair in terms of achievable rate and peak power is presented.  
\section{System Model}
\begin{figure}[h!]
  \begin{center}
    \includegraphics[width=\columnwidth, draft=false, trim=0 14mm 0 21mm, clip=true]{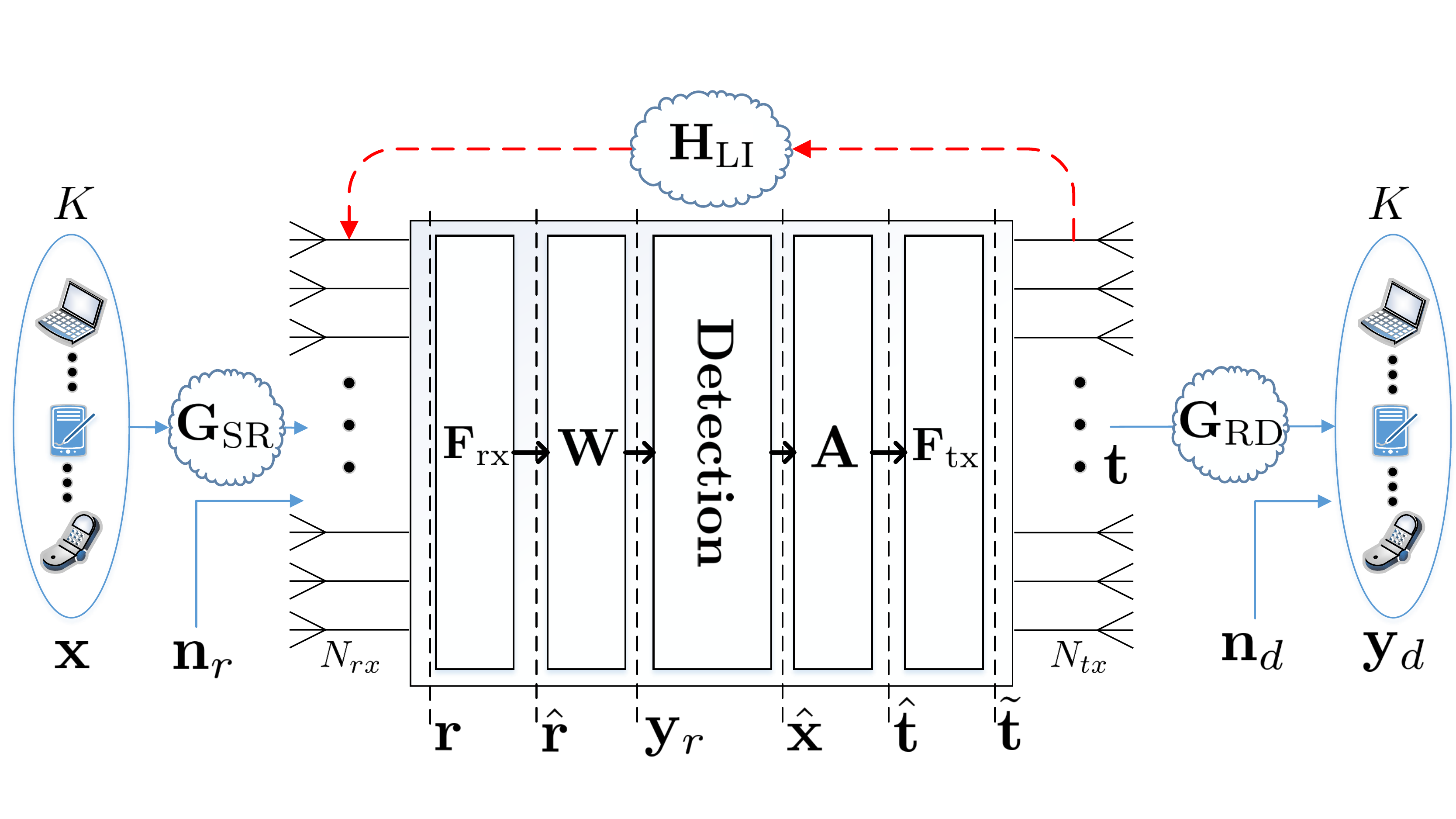}
  \end{center}
  \caption{Full-duplex relay system.}
  \label{fig:relay}
\end{figure}

Let us consider $K$ user pairs establish a wireless connection through a relay station, which operates in in-band full-duplex mode (sharing the same time-frequency resources). For this purpose, each user is equipped with a single antenna, while the relay has $N_{rx}$ receiving antennas and $N_{tx}$ transmitting antennas. The number of antennas at the relay may go up to a few hundred and the number of served users is in general considered to be much lower ($K<<N_{rx},N_{tx}$). The nonexistence of a direct link between the sources and destinations is assumed, hence all links are established via the relay station. A simplified representation, as well as the used notation to be defined in this section, can be seen in figure \ref{fig:relay}.

\label{sec:sd}

\subsection{Channel Model}

\looseness 0 For the evaluation of the performance of such setup, we are interested in the vectors containing the transmitted symbols from the sources to the relay and the beamformed symbols from the relay to the destinations, denoted by $\mathbf{x} \in \mathbb{C}^{K\times 1}$ and $\mathbf{t} \in \mathbb{C}^{N_{tx}\times 1}$, respectively. These vectors are assumed to be taken from an $M$-QAM constellation and the output signals to be normalized and uncorrelated, such that $\mathbb{E}(\mathbf{x}\mathbf{x}^H)=\mathbf{I} $ and $\mathbb{E}(\mathbf{t}^H\mathbf{t})=1$. 
In this manner, the power transmitted by each source is independent of $K$ and the total transmitted power at the relay dissociated of $N_{tx}$. The received vectors at the relay and at the destinations are given by \eqref{eq:signalrelay} and \eqref{eq:signaldest}, respectively. 
\begin{align}
&\mathbf{r}=\mathbf{G}_{\text{SR}}\mathbf{D}_{p_\text{S}}^{1/2}\mathbf{x}+\sqrt{p_\text{R}}\mathbf{H}_\text{LI}\mathbf{t}+\mathbf{n}_r. \label{eq:signalrelay}\\
&\mathbf{y}_d=\sqrt{p_\text{R}}\mathbf{G}_\text{RD}\mathbf{t}+\mathbf{n}_d. \label{eq:signaldest}
\end{align}   
  
\looseness 0 The uplink symbols in $\mathbf{x}$ pass through a MIMO channel from the sources to the relay denoted by $\mathbf{G}_{\text{SR}} \in \mathbb{C}^{N_{rx} \times K}$, whereas the symbols from the relay to the destinations pass through $\mathbf{G}_{\text{RD}} \in \mathbb{C}^{K \times N_{tx}}$. Both matrices are expressed as $\mathbf{G}_{\text{SR}} = \mathbf{H}_\text{SR}\mathbf{D}_{\text{SR}}^{1/2}$ and $\mathbf{G}_{\text{RD}} = \mathbf{D}_{\text{RD}}^{1/2}\mathbf{H}_{\text{RD}}$, where $\mathbf{D}_{\text{SR}}$ and $\mathbf{D}_{\text{RD}}$ are diagonal matrices with entries $\beta_{\text{SR,}k}$ and $\beta_{\text{RD,}k}$ taken from a log-normal distribution, which account for large-scale channel effects. The fast-fading channel components are present in $\mathbf{H}_{\text{SR}}$ and $\mathbf{H}_{\text{RD}}$, both with independent entries taken from a $\mathcal{CN}(0,1)$ distribution. The LI channel is represented by $\mathbf{H}_{\text{LI}}$ with distribution $\mathcal{CN}(0,\sigma_{\text{LI}}^2)$, where $\sigma_{\text{LI}}^2$ accounts for the residual LI power after self-interference suppression imposed in the analog-domain. The vectors $\mathbf{n}_r \sim \mathcal{CN}(0,\sigma_{\mathbf{n}_r}^2)$ and $\mathbf{n}_d \sim \mathcal{CN}(0,\sigma_{\mathbf{n}_d}^2)$ account for additive white Gaussian noise  at the relay and destinations, respectively. Finally, the diagonal matrix $\mathbf{D}_{p_\text{S}}$ with entries $p_\text{S,k}$ regulates each source transmit power, while $p_\text{R}$ denotes the relay's average transmit power.

\subsection{Channel Estimation}

In order to efficiently apply  detection, precoding and LI mitigation techniques, channel estimates $\mathbf{\tilde{H}}_{\text{SR}}$, $\mathbf{\tilde{H}}_{\text{LI}}$ and $\mathbf{\tilde{H}}_{\text{RD}}$ of the true $\mathbf{H}_{\text{SR}}$, $\mathbf{H}_{\text{LI}}$ and $\mathbf{H}_{\text{RD}}$ are required. Applying any of the estimation algorithms proposed in the literature \cite{mel:fdwireles}, we assume an error in the channel matrices modelled by $\mathcal{E}_{\mathbf{H}_{\text{SR}}}, \mathcal{E}_{\mathbf{H}_{\text{RD}}}, \mathcal{E}_{\mathbf{H}_{\text{LI}}} \sim \mathcal{CN}(0,\epsilon^2_{\mathbf{H}})$. The difference between the non-ideal estimates and the true channel values are
\begin{equation}
    \label{eq:errormatrix}
    \mathbf{H}_{\text{SR}} = \mathbf{\tilde{H}}_{\text{SR}} + \mathcal{E}_{\mathbf{H}_{\text{SR}}};\\  
    \mathbf{H}_{\text{LI}} = \mathbf{\tilde{H}}_{\text{LI}} + \mathcal{E}_{\mathbf{H}_{\text{LI}}};\\
    \mathbf{H}_{\text{RD}} = \mathbf{\tilde{H}}_{\text{RD}} + \mathcal{E}_{\mathbf{H}_{\text{RD}}}.  
    \end{equation}
No errors in the large-scale fading matrices are assumed. The impact of any hardware imperfection 
at the relay is modeled by means of an additive error \cite{riihonen2011mitigation} in the transmitted vector given by $\mathbf{t} = \tilde{\mathbf{t}} + \mathcal{E}_{\mathbf{t}}$, where $\tilde{\mathbf{t}}$ is the vector to be transmitted after baseband filtering and where all the elements of \mbox{$\mathcal{E}_{\mathbf{t}}\sim\mathcal{CN}(0,\epsilon^2_{\mathbf{t}})$} are  assumed to be uncorrelated with $\tilde{\mathbf{t}}$. It is worth mentioning that the covariance of the LI term $\sqrt{p_\text{R}}\mathbf{H}_{\text{LI}}\mathbf{t}$ in (\ref{eq:signalrelay}) is controlled by the parameters $p_\text{R}$, $\sigma_{\text{LI}}^2$, $\epsilon_{\mathbf{t}}^2$ and $\epsilon_{\mathbf{H}}^2$.

\subsection{Detection and Precoding}
Let us consider for the moment an equivalent received signal $\hat{\mathbf{r}}$, whose LI component has been minimized. Hence, we are in the presence of an uplink channel where $K<<N_{rx}$. Under this assumption, linear processing techniques are proven to perform close to optimal \cite{hien-multi}. We consider the usage of zero-forcing (ZF) filtering for both detection and beamforming \cite{yang2013performance}. Using the estimation of $\mathbf{G}_{\text{SR}}$, denoted by $\tilde{\mathbf{G}}_{\text{SR}} = \tilde{\mathbf{H}}_{\text{SR}}\mathbf{D}^{1/2}_{\text{SR}} $, the estimated symbols after a ZF filter $\mathbf{W}_{\text{zf}}$ are given by    
\begin{equation}
\hat{\mathbf{x}} = \mathcal{Q}(\mathbf{W}_{\text{zf}}\hat{\mathbf{r}}) = \mathcal{Q}\big((\tilde{\mathbf{G}}_{\text{SR}}^H\tilde{\mathbf{G}}_{\text{SR}})^{-1}\tilde{\mathbf{G}}_{\text{SR}}^H\hat{\mathbf{r}}\big),
\end{equation}
where $\mathcal{Q}(\cdot)$ is a symbol-wise quantizer to the constellation set.
Upon detection based on $\hat{\mathbf{x}}$, the estimated symbols are forwarded to the destinations $\mathbf{y}_d$, which we assume to have very limited processing capabilities. Thus, a ZF precoder $\mathbf{A}_{\text{zf}}$ is used. The corresponding expression is 
\begin{equation}
\hat{\mathbf{t}} = \mathbf{A}_{\text{zf}}\hat{\mathbf{x}} = \alpha_{\text{zf}} \tilde{\mathbf{G}}_{\text{RD}}^H (\tilde{\mathbf{G}}_{\text{RD}}\tilde{\mathbf{G}}_{\text{RD}}^H)^{-1}{\mathbf{\hat{\mathbf{x}}}},
\end{equation}
where $\tilde{\mathbf{G}}_{\text{RD}}=\mathbf{D}^{1/2}_{\text{RD}} \tilde{\mathbf{H}}_{\text{RD}}$ is the estimation of the true $\mathbf{G}_{\text{RD}}$ and $\alpha_{\text{zf}} = \big(\mathbb{E}(tr\{(\tilde{\mathbf{G}}_{\text{RD}}\tilde{\mathbf{G}}_{\text{RD}}^H)^{-1}\})\big)^{-\frac{1}{2}}$ is a scalar chosen to normalize the power of $\hat{\mathbf{t}}$, i.e., $\mathbb{E}(\hat{\mathbf{t}}^H\hat{\mathbf{t}})=1$. From the definition of $\tilde{\mathbf{H}}_{\text{RD}} \sim \mathcal{CN}(0,1+\epsilon_{\mathbf{H}}^2)$, $\alpha_{\text{zf}}$ comes as (cf. appendix A)
\begin{equation}
\label{eq:alfazf}
\alpha_{\text{zf}} = \sqrt{\frac{(N_{tx}-K)}{\sum_{k=1}^K \big(\beta_{\text{RD,}k}(1+\epsilon_{\mathbf{H}}^2)\big)^{-1}}}.
\end{equation}

\section{Loopback Interference Mitigation}

\subsection{Linear Filtering}
\looseness -1 We now evaluate  the problem of suppressing LI, using linear processing. Under the model given by (\ref{eq:signalrelay}), we are interested in minimizing the LI term $\sqrt{p_\text{R}}\mathbf{H}_{\text{LI}}\mathbf{t}$, while preserving the signal $\mathbf{G}_{\text{SR}}\mathbf{D}_{p_\text{S}}^{1/2}\mathbf{x}$ and taking into account the covariance of the noise vector $\mathbf{R}_{\mathbf{n}_r}$ at the relay. Hence, the mean square error between the desired signal and the one received by the relay station is $\mathbb{E}((\mathbf{G}_\text{SR}\mathbf{D}_{p_\text{S}}^{1/2}\mathbf{x}-\hat{\mathbf{r}})(\mathbf{G}_\text{SR}\mathbf{D}_{p_\text{S}}^{1/2}\mathbf{x}-\hat{\mathbf{r}})^H)$. Using the method proposed in \cite{riihonen2011mitigation}, we consider a linear prefilter $\mathbf{F}_{\text{tx}}$ and postfilter $\mathbf{F}_{\text{rx}}$, such that $\tilde{\mathbf{t}}=\mathbf{F}_{\text{tx}}\hat{\mathbf{t}}$ and $\hat{\mathbf{r}}=\mathbf{F}_{\text{rx}} \mathbf{r}$. The filter that finds the MMSE when $\mathbf{F}_{\text{rx}}$ is fixed satisfies (cf. appendix B)
\begin{equation}
\label{nsp}
\mathbf{F}_{\text{rx}} \tilde{\mathbf{H}}_{\text{LI}} \mathbf{F}_{\text{tx}} = \mathbf{0},
\end{equation}
and when $\mathbf{F}_{\text{tx}}$ is fixed becomes
\begin{equation}
\label{mmsefilter}
\mathbf{F}_{\text{rx}} = \tilde{\mathbf{G}}_{\text{SR}} \mathbf{D}_{p_\text{S}} \tilde{\mathbf{G}}_{\text{SR}}^H (\tilde{\mathbf{G}}_{\text{SR}} \mathbf{D}_{p_\text{S}} \tilde{\mathbf{G}}_{\text{SR}}^H + p_\text{R}\tilde{\mathbf{H}}_{\text{LI}} \mathbf{R}_\mathbf{t}\tilde{\mathbf{H}}_{\text{LI}}^H + \mathbf{R}_{\mathbf{n}_r})^{-1},
\end{equation}
\looseness 0 where $\mathbf{R}_\mathbf{t}=\mathbf{F}_{\text{tx}}\mathbf{A}_{\text{zf}}\mathbf{R}_{\hat{\mathbf{x}}}\mathbf{A}_{\text{zf}}^{H}\mathbf{F}_{\text{tx}}^{H}+\epsilon^2_{\mathbf{t}}\mathbf{I}$. The solution provided by (\ref{nsp}) corresponds to the null-space projection already presented in \cite{riihonen2011mitigation}, and in general finding a reliable solution for large dimensions is hard. On the contrary, \eqref{mmsefilter} is a closed-form expression containing the channel estimates and covariance matrices of transmitted vectors and noise, that are assumed to be known.
Under the considered model and for a given set of parameters, a trade-off in the e2e BER is expected: on the one hand, a lower transmitted power at the relay $p_\text{R}$ reduces the LI power and diminishes any impact of the estimation errors and radio-frequency impairments; on the other hand, a higher $p_\text{R}$ leads to improved signal-to-noise ratios (SNR) levels at the destinations and hence a lower BER in the forward link channel. Thus, 
for a fixed source transmit power, an optimal choice for $p_R$ that minimizes the e2e BER can be anticipated. 

\subsection{Optimal Power Allocation}
\label{sec:opa}	
The level of interference suffered by the relay station depends critically on its transmission power and directly perturbs the e2e performance. Thus, finding the optimal power that meets the requirements of the system is desired. Let us consider the achievable rate for each individual link. The procedure done in \cite{riihonen2011power} allows us to obtain an expression for the transmission link rate between each source and destination
\begin{equation}
\label{eq:ratemin}
R_k=\min\{ R_{\text{SR,}k}, R_{\text{RD,}k}\},
\end{equation}
where $R_{\text{SR,}k}$ and $R_{\text{RD,}k}$ denote the achievable rates between the sources and the relay and between the relay and the destinations, respectively. 
To obtain $R_k$, 
we start by considering the received signal at the relay before detection given by
\begin{equation}
\label{eq:relayrec}
\begin{split}
y_{\text{r,}k}=&\sqrt{p_{\text{S,}k}}(\mathbf{W}_\text{zf}\mathbf{F}_\text{rx})_k^T \mathbf{g}_{\text{SR},k} x_k\\ 
&+\sum\nolimits_{j \neq k}^K \sqrt{p_{\text{S,}j}}(\mathbf{W}_\text{zf}\mathbf{F}_\text{rx})_k^T \mathbf{g}_{\text{SR},j} x_j  \\
&+\sqrt{p_\text{R}}(\mathbf{W}_\text{zf}\mathbf{F}_\text{rx})_k^T \mathbf{H}_\text{LI}\mathbf{t}+ (\mathbf{W}_\text{zf}\mathbf{F}_\text{rx})_k^T \mathbf{n}_r, 
\end{split}
\end{equation}
where $\mathbf{g}_{\text{SR},k}$ denotes the $k^{th}$ column of $\mathbf{G}_{\text{SR}}$. The received signal at each destination link before detection is given by
\begin{equation}
\label{eq:destrec}
\begin{split}
y_{\text{d,}k}=&\sqrt{p_\text{R}} { \mathbf{g}^T_{\text{RD},k} } ( \mathbf{F}_\text{tx} \mathbf{A}_\text{zf} )_k \hat{x}_k\\
&+\sqrt{p_\text{R}}\sum\nolimits_{j \neq k}^K  { \mathbf{g}^T_{\text{RD},k} } ( \mathbf{F}_\text{tx} \mathbf{A}_\text{zf} )_j  \hat{x}_j + n_{d,k}.
\end{split}
\end{equation}
\looseness 0 Both expressions may be seen as a known mean gain times the desired signal plus an uncorrelated effective noise term that includes channel impairment effects, interpair and loopback interference, and Gaussian noise. A 
technique commonly used in large MIMO systems \cite{hassibi2003} is to approximate the effective noise term by a Gaussian noise term under the central limit theorem, and which gives a good approximation for $R_{\text{SR,}k}$ as in \eqref{eq:rg_sr} and for $R_{\text{RD,}k}$ as in \eqref{eq:rg_rd}.

The goal is to find the system power allocation, i.e., compute the required power transmitted by both sources and relay, such that the desired rate for each communication pair $k$ is guaranteed. Moreover, peak power constraints should be satisfied and overall power consumption taken into consideration; in other words, the system energy efficiency, defined as \mbox{$\text{EE}=\sum_{k=1}^K R_{k} /(p_{\text{R}} + \sum_{k=1}^K p_{\text{S,}k})$}, should be maximized. This can be described as the following optimization problem  
\begin{equation}
\label{eq:optproblem}
\begin{aligned}
& \underset{ p_{\text{R}} , {p}_{\text{S},1}... {p}_{\text{S},K}  }{\text{min.}}
& & \sum\nolimits_{k=1}^{K} p_{\text{S,}k}+p_\text{R}, \\
& \text{s.t.}
&& {\scriptstyle R_{k}  \geq R_{0,k}, \; k = 1, \ldots, K;}\\ 
&&&{\scriptstyle 0 \le p_{\text{S,}k} \le p_{\text{S}_{0,k}}, \; k = 1, \ldots, K;}\\
&&&{\scriptstyle 0 \le p_{\text{R}} \le p_{\text{R}_{0}},}
\end{aligned}
\end{equation} 
where $R_{0,k}$ and $p_{\text{S}_{0,k}}$ are the required rate and peak power for pair $k$ respectively, and $p_{\text{R}_\text{0}}$ is the relay station peak power.
\begin{figure*}[]
\begin{equation}
\label{eq:rg_sr}
R_{\text{SR,}k} = \log_2 \Big ( 1 + {\frac{ p_{\text{S,}k} \text{MV}_{\text{SR,}k}     }{  p_{\text{S,}k} \text{V}_{\text{SR,}k} + \sum_{ 
j \neq k}^{K} p_{\text{S,}j} \text{MP}_{\text{SR,}(k,j)} + p_{\text{R}} \text{LI}_{\text{SR,}k} +  \text{AN}_{\text{SR,}k}  }}  \Big ), 
\end{equation}
where $\text{MV}_{\text{SR,}k}=| \mathbb{E} \{ (\mathbf{W}_\text{zf}\mathbf{F}_\text{rx})_k^T \mathbf{g}_{\text{SR},k} \}|^2,$
$\text{V}_{\text{SR,}k}=\mathbb{V}\text{ar}\{(\mathbf{W}_\text{zf}\mathbf{F}_\text{rx})_k^T \mathbf{g}_{\text{SR},k}\},$
$\text{MP}_{\text{SR,}(k,j)}= \mathbb{E} \{ | (\mathbf{W}_\text{zf}\mathbf{F}_\text{rx})_k^T \mathbf{g}_{\text{SR},j} |^2 \},$ 

$\text{LI}_{\text{SR,}k}=  \mathbb{E} \{ \parallel \mathbf{w}_{\text{zf},k}^T \mathbf{F}_\text{rx}  \mathbf{H}_\text{LI} \mathbf{F}_\text{tx} \mathbf{A}_\text{zf} \parallel^2 \} \text{ and}$ 
 $\text{AN}_{\text{SR,}k}=\sigma_{n_r}^2  \mathbb{E} \{\parallel(\mathbf{W}_\text{zf}\mathbf{F}_\text{rx})_k  \parallel^2 \}.$

\begin{equation}
\label{eq:rg_rd}
 R_{\text{RD,}k} = \log_2 \Big ( 1 +  { \frac{ p_{\text{R}} \text{MV}_{\text{RD,}k} }{  p_{\text{R}} \text{V}_{\text{RD,}k}  + p_{\text{R}} \text{MP}_{\text{RD,}k} + \text{AN}_{\text{RD,}k} } }  \Big ),
\end{equation}
where \resizebox{0.95\hsize}{!}{$\text{MV}_{\text{RD,}k}=| \mathbb{E} \{ {\mathbf{g}_{\text{RD},k}^T} (\mathbf{F}_\text{tx}  \mathbf{A}_{\text{zf}})_{k} \} |^2,$
$\text{V}_{\text{RD,}k}= \mathbb{V}\text{ar}\{\mathbf{g}_{\text{RD},k}^T (\mathbf{F}_\text{tx}  \mathbf{A}_{\text{zf}})_{k}\},
\text{MP}_{\text{RD,}k}=\sum\limits_{j \neq k}^{K}  \mathbb{E} \{ | {\mathbf{g}_{\text{RD},k}^T} (\mathbf{F}_\text{tx}  \mathbf{A}_{\text{zf}})_{j} |^2 \}\text{ and }\text{AN}_{\text{RD,}k}=\sigma_{n_d}^2.$} 
\vspace{0.05cm}
\hrule
\end{figure*}
The solution to problem \eqref{eq:optproblem} involves the computation of the channel statistics which are the output of involved expressions due to the non-linear dependence of the optimization variables in $\mathbf{F}_\text{rx}$. For that reason, we propose a simple and computationally efficient iterative algorithm, based on linear programming, capable of obtaining the referred OPA. The algorithm computes the channel statistics coefficients in \eqref{eq:rg_sr} and \eqref{eq:rg_rd}, the loopback mitigation filters and the power vectors iteratively until the optimal power vector is reached. The procedure is summarized in algorithm \ref{alg1}.
\begin{algorithm}[h]
\setstretch{1.2}
\caption{}
\label{alg1}
\textbf{1. Initialization: }  Set $i=1$; initialize powers $p_{\text{S},k,1}=p_{\text{S}_{0,k}}$ and $p_{\text{R,0}}=p_{\text{R}_\text{0}}$; define $L$ as the total number of iterations and set $N_{it}$ as the number of channel realizations per iteration.  

\textbf{2. Iteration i: }

\begin{algorithmic}

\STATE 1) Compute channel statistics:
	\FOR {$N_{it}$ } 
	
	\STATE i) 	Generate $\mathbf{G}_\text{SR}$, $\mathbf{G}_\text{RD}$, $\mathbf{H}_\text{LI}$, $\mathbf{W}_\text{zf}$ and $\mathbf{A}_\text{zf}$
	\STATE ii) 	Compute filter $\mathbf{F}_\text{rx}$ with $p_{\text{S,}k,i}$ and $p_{\text{R,}i}$   
	\STATE iii) Compute instantaneous rate  coefficients for all $k$ pairs:
		 $\text{MV}_{\text{SR,}k}$, $\text{V}_{\text{SR,}k}$, $\text{MP}_{\text{SR,}(k,j)}$, $\text{LI}_{\text{SR,}k}$ and $\text{AN}_{\text{SR,}k}$ (as in \eqref{eq:rg_sr})
		 $\text{MV}_{\text{RD,}k}$, $\text{V}_{\text{RD,}k}$, $\text{MP}_{\text{RD,}k}$ and $\text{AN}_{\text{RD,}k}$ (as in \eqref{eq:rg_rd})  
	\ENDFOR
	
\STATE 2) Average to obtain channel statistics.
\STATE 3) Solve the linear program \eqref{eq:optproblem} with the coefficients found in {step~2)} to obtain the new $p_{\text{S,}k,i}$ and $p_{\text{R,}i}$. 

\STATE 4) Set $p_{\text{S},k,i+1}=p_{\text{S,}k,i}$ and $p_{\text{R,}i+1}=p_{\text{R,}i}$.

\end{algorithmic}

\textbf{3. Check}: If $i=L$ end algorithm, else set $i=i+1$.
\end{algorithm}

\section{Numerical Results}
\label{sec:nr}

In this section, we compare the performance of the proposed filters with special emphasis on the e2e link reliability and on the efficiency of the power allocation scheme. All results are obtained via Monte Carlo simulation, using uncoded MIMO. 

\subsection{System Parameters}

We assume a symmetric system with the same number of users $K$ on both sides with one antenna each and $N = N_{tx} = N_{rx}$ antennas at the relay. Transmission and reception at the relay are done in the same time-slot and frequency band, and an arbitrary processing delay $d \geq 1$ is assumed such that at time instant $i$: $\hat{\mathbf{x}}[i] = f(\mathbf{x}[i-d])$. We consider $\mathbf{F}_\text{rx}$ as in \eqref{mmsefilter} and $\mathbf{F}_\text{tx}=\mathbf{I}$. Without loss of generality, we also consider $\sigma_\text{LI}^2=1$. The SNR at the relay is defined as $\text{SNR}_\text{R} = \frac{\sum_{k=1}^K \beta_{\text{SR,}k} {p}_{\text{S},k}}{\sigma^2_{\mathbf{n}_r}}$. 


\subsection{Transmitted Power vs. BER}
\label{sec:nrber}
\begin{figure}[t]
  \begin{center}
    \includegraphics[width=\columnwidth, trim=0 2mm 0 8mm, clip=true, draft=false]{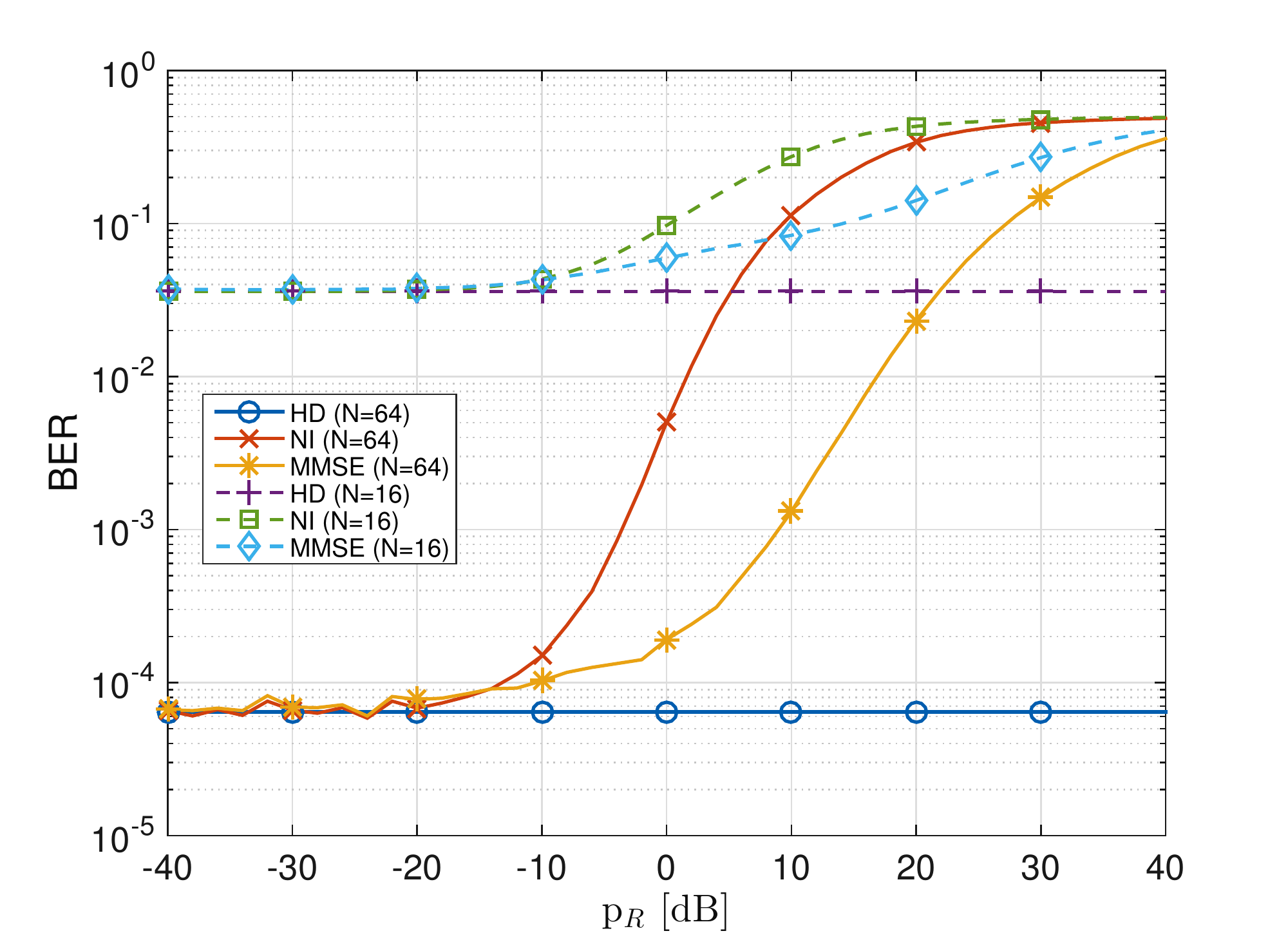}
  \end{center}
  \vspace{-3mm}
  \caption{BER performance at the relay for different numbers of antennas $N$, $K=5$ pairs, $\text{SNR}_\text{R}=8$ dB and 16-QAM.}
  \label{fig:bervsli}
\end{figure}

\begin{figure}[t]
  \begin{center}
    \includegraphics[width=\columnwidth, trim=0 2mm 0 8mm, clip=true, draft=false]{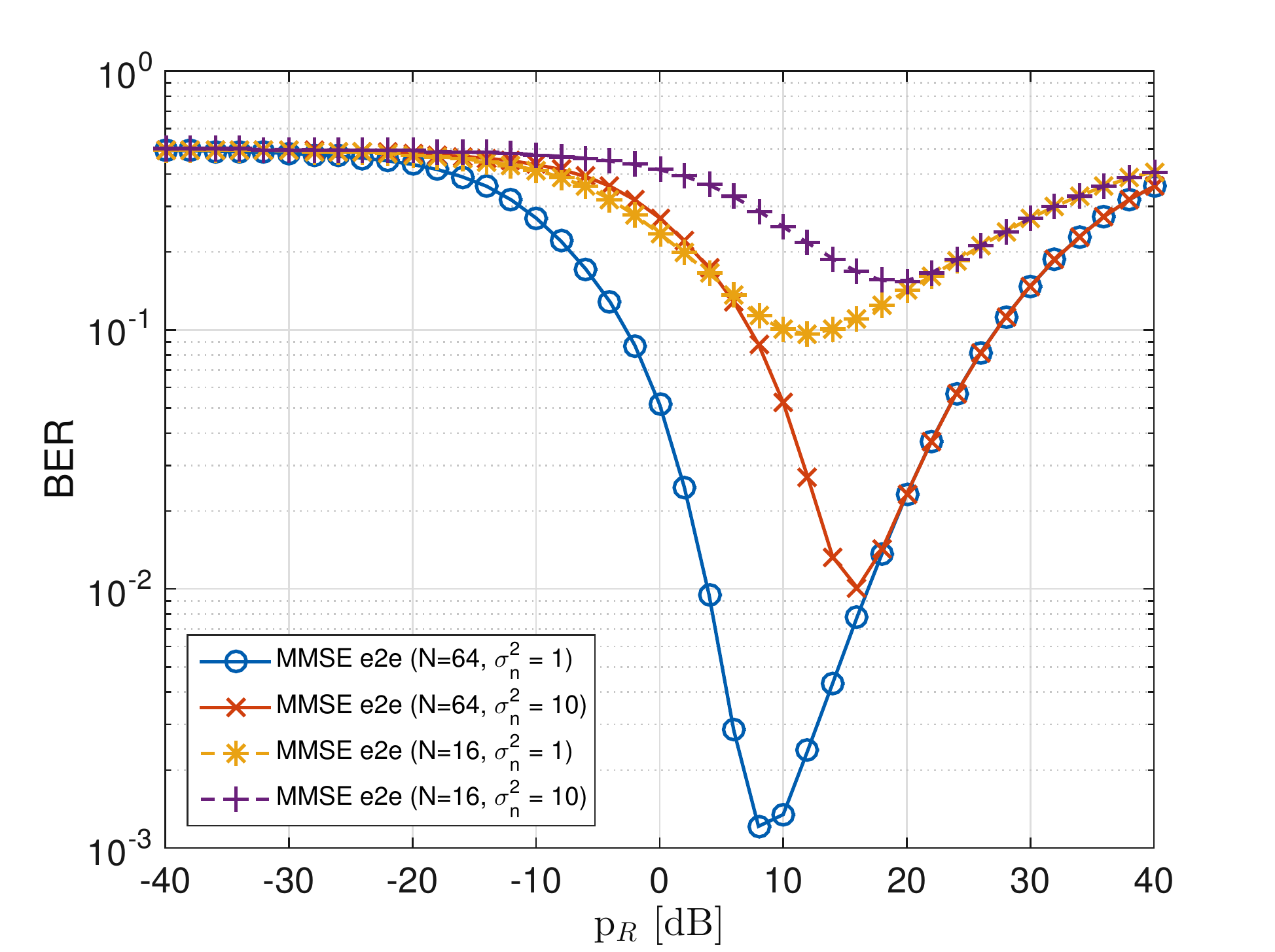}
  \end{center}
  \vspace{-3mm}
  \caption{End-to-end BER performance for different $N$ and $\sigma^2_{\mathbf{n}_d}$, $K=5$ pairs, $\text{SNR}_\text{R}=8$ dB and 16-QAM.} \vspace{-4mm}
  \label{fig:bere2evsli}
\end{figure}

We start by evaluating the performance of the system in terms of BER at both the relay and destinations, considering only small-scale fading, that is $\mathbf{D}_\text{SR}^{1/2} = \mathbf{D}_\text{RD}^{1/2} = \mathbf{I}$. Firstly, for a given $\text{SNR}_\text{R}$ and fixed uniform transmitted power $p_\text{S,k}=1$ for all $k$, we are interested in studying different allocated powers at the relay $p_\text{R}$ and for an increasing number of antennas $N$. For comparison purposes, we implemented natural isolation (NI) (when $\mathbf{F}_\text{rx}=\mathbf{F}_\text{tx}=\mathbf{I}$, which corresponds to ignoring the LI component) and the HD counterpart (when $\mathbf{t}=0$ and $\mathbf{F}_\text{rx}=\mathbf{F}_\text{tx}=\mathbf{I}$). Setting the variance in the errors to be $\epsilon^2_{\mathbf{t}}=\epsilon^2_{\mathbf{H}}=10^{-3}$, curves of BER for different numbers of antennas and relay power $p_\text{R}$ are depicted in figure \ref{fig:bervsli}. It can be seen that for an increasing number of antennas, the system using the MMSE filter becomes more robust to the LI, showing a BER performance closer to the HD for higher values of power $p_\text{R}$ than when compared with NI. 
We then evaluate the e2e BER in terms of the allocated power $p_\text{R}$ and using the same setup as before. For this purpose, we set the variance of the noise at the destinations $\sigma^2_{\mathbf{n}_d}$, quantize the received signals in \eqref{eq:signaldest} and compare them with the symbols transmitted by the sources. The results are shown in figure \ref{fig:bere2evsli}. One can confirm that for a given configuration there is an optimal choice for the power at the relay that minimizes the e2e BER. Moreover, it is concluded that a larger number of antennas attains the minimum BER with less power. 

%
%
%
\subsection{Optimal Power Allocation Algorithm}
\label{sec:opaAl}

\begin{figure}[t]
  \begin{center}
    \includegraphics[width=\columnwidth, trim=0 3mm 0 8mm, clip=true, draft=false]{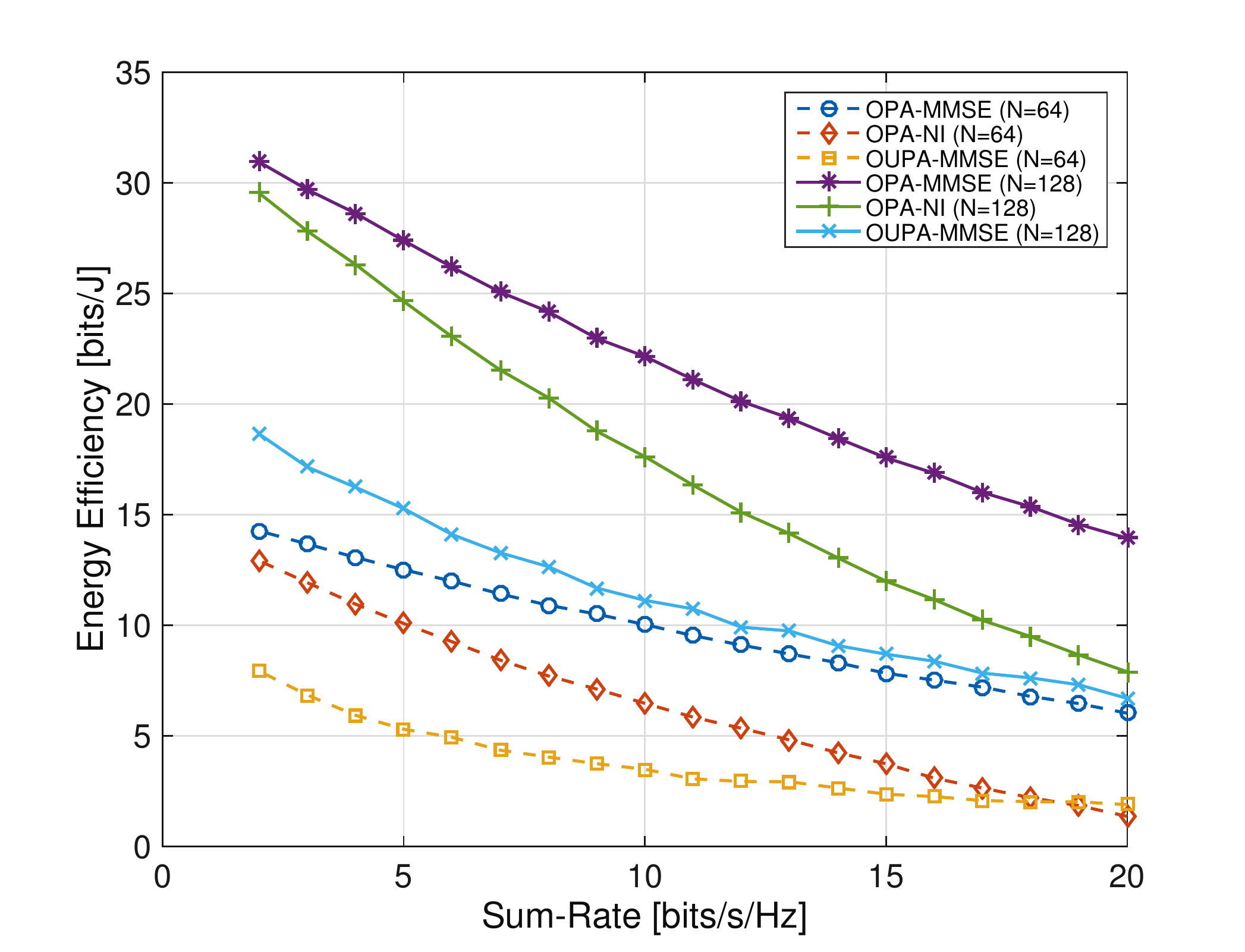}
  \end{center}
  \vspace{-3mm}
  \caption{Energy efficiency for different power allocation schemes, filters and number of antennas. The remaining parameters are $\sigma^2_{\mathbf{n}_d}=1$, $K=10$ pairs and $\text{SNR}_\text{R}=16$ dB.} \vspace{-4mm}
  \label{fig:alg_comp}
\end{figure}

Finally, we evaluate the results of the proposed algorithm in section \ref{sec:opa}, finding the OPA that meets the  rate constraints of each link, while minimizing the overall power of the system. Large-scale fading effects are now taken into consideration, more precisely, $\beta_{\text{SR,}k}$ and $\beta_{\text{RD,}k}$ are independent variables generated from a log-normal distribution with mean value $m=1$ and standard deviation $\sigma=6$ dB. 
Additionally, we set the normalized peak power of the sources and relay as $p_{\text{S}_{0,k}}=3$ dB and $p_{\text{R}_\text{0}}=10$ dB, respectively. In addition, parameters $N_{it}=10^3$ and $L=5$ are empirically defined, such that both channel statistics and output powers in algorithm \ref{alg1} are good approximations of their real value. The performance of algorithm \ref{alg1} is determined in terms of EE for the case where interference effects are disregarded (OPA-NI) and when the MMSE filter is used (OPA-MMSE). This method is also compared with the situation where an optimal uniform power allocation (OUPA) is used, which corresponds to consider $p_{\text{S},k}=p_\text{S}$, for all $k$, in \eqref{eq:optproblem}. Figure \ref{fig:alg_comp} shows the  curves of average EE for different values of desired e2e sum-rate, defined as $\sum_{k=1}^{K} R_{0,k}$, where the individual required rates $R_{0,k}$ are taken from a discrete uniform distribution.    

As one may see, for the same desired sum-rate, the EE of OPA-MMSE is improved significantly when compared to OUPA, while guaranteeing that no link is in outage \mbox{($R_{k} \geq R_{0,k}$} for all $k$). Furthermore, since MMSE filter effectively reduces the LI, a lower amount of energy is spent to achieve the same sum-rate when compared with OPA-NI, an effect that becomes more preponderant in higher rates regimes.

\section{Conclusions}
\label{sec:conclusions}

\looseness -1 This paper studies an independent single-antenna multipair communication system using an in-band full-duplex relay station equipped with a large number of antennas. Considering imperfect channel state information, the relay applies two stages of filtering. Firstly, inner ZF filters are used in order to perform both detection and beamforming. Secondly, an MMSE outer filter is designed to mitigate LI, while preserving the e2e channel. We assess the effectiveness of the MMSE filter via simulation in terms of BER at the relay and comparing it with NI and the HD counterpart. Moreover, in the e2e BER analysis, an optimal choice for the power at the relay was identified. We further confirmed the low complexity, high performance capabilities of linear processing in systems employing massive MIMO. We provide the expressions for the e2e achievable rate of the studied setup and propose a low complexity iterative algorithm to perform OPA. The latter takes into consideration large-scale fading effects, peak power constraints and the rate requirements for each individual pair. The results were evaluated in terms of EE, and the system employing MMSE filtering outperformed both NI and OUPA, specially for high sum-rates.

\section*{Appendix}
\subsection{Normalization factor $\alpha_{\text{zf}}$}
\label{ap:azf}
The objective is to design $\alpha_\text{zf}$ such that $\mathbb{E}\{\mathbf{t}^H\mathbf{t}\}=1$. Using its definition $\mathbf{t} = \alpha_\text{zf}\tilde{\mathbf{G}}_\text{RD}^H(\tilde{\mathbf{G}}_\text{RD}
\tilde{\mathbf{G}}_\text{RD}^H)^{-1}\hat{\mathbf{x}}=
\alpha_\text{zf}\mathbf{P}\hat{\mathbf{x}}$,
and the fact that $\mathbf{C}=\mathbb{E}\{\hat{\mathbf{x}}\hat{\mathbf{x}}^H\}=\mathbf{I}$, it follows that 
\vspace{0mm}
\begin{equation}
\label{eq:normcov}
\mathbb{E}\{\mathbf{t}^H\mathbf{t}\} = \alpha_\text{zf}^2 \mathbb{E} \{ (\mathbf{P}\hat{\mathbf{x}})^H (\mathbf{P}\hat{\mathbf{x}})\} = \alpha_\text{zf}^2 tr\{\mathbf{PC}\mathbf{P}^H\} = 1.  
\end{equation}  
Exploiting $\mathbf{C} = \mathbf{I}$, it follows that $tr\{\mathbf{PCP}^H\}$ is given by
$
	tr\{ (\tilde{\mathbf{G}}_\text{RD}\tilde{\mathbf{G}}_\text{RD}^H)^{-1}\}= tr\{(\mathbf{D}_\text{RD} \tilde{\mathbf{H}}_\text{RD} \tilde{\mathbf{H}}_\text{RD}^H)^{-1}\}.$
Now noting that $\mathbf{W} = \tilde{\mathbf{H}}_\text{RD} \tilde{\mathbf{H}}_{RD}^H$ is a central Wishart matrix, where the columns of $\tilde{\mathbf{H}}_\text{RD} \in \mathbb{C}^{K\times N_{tx}}$ are zero-mean complex Gaussian vectors with covariance matrix $(1+\epsilon_\mathbf{H}^2)\mathbf{I}$, it follows from \cite[Lemma 2.10]{tulinothelegend} that
\begin{equation}
\label{eq:wishart}
\mathbb{E}\{ (tr\{\mathbf{D}_\text{RD}\mathbf{W} \} )^{-1} \} = \frac{\sum_{k=1}^K \big(\beta_\text{RD,k}(1+\epsilon_{\mathbf{H}}^2)\big)^{-1}}{N_{tx}-K}.
\end{equation}

Merging 
(\ref{eq:normcov}) and (\ref{eq:wishart}), the 
factor $\alpha_\text{zf}$ in (\ref{eq:alfazf}) follows.

\subsection{$\mathbf{F}_{tx}$ and $\mathbf{F}_{rx}$ expressions}
\label{ap:gtxgrx}
The error covariance matrix $\mathbf{Q}$ of the relay input signal is given by $\mathbb{E}((\mathbf{G}_\text{SR}\mathbf{x}-\hat{\mathbf{r}})(\mathbf{G}_\text{SR}\mathbf{x}-\hat{\mathbf{r}})^H)$, yielding \cite{riihonen2011mitigation}
\begin{equation}
\begin{split} 
\mathbf{Q} =& (\mathbf{I}-\mathbf{F}_\text{rx})\tilde{\mathbf{G}}_\text{SR}\mathbf{D}_{p_\text{S}} \tilde{\mathbf{G}}_\text{SR}^H (\mathbf{I}-\mathbf{F}_\text{rx})^H \\ 
&+p_\text{R} \mathbf{F}_\text{rx} \tilde{\mathbf{H}}_\text{LI}\mathbf{R}_t \tilde{\mathbf{H}}_\text{LI}^H \mathbf{F}_\text{rx}^H +\mathbf{F}_\text{rx}\mathbf{R}_{\mathbf{n}_r} \mathbf{F}_\text{rx}^H,
 \end{split}
\end{equation}
where $\mathbf{R}_{\mathbf{t}} = \mathbf{R}_{\tilde{\mathbf{t}}} + \mathbf{R}_{\mathcal{E}_\mathbf{t}}=\mathbb{E}(\tilde{\mathbf{t}}\tilde{\mathbf{t}}^H)+\mathbb{E}(\mathcal{E}_\mathbf{t}\mathcal{E}_\mathbf{t}^H)=\mathbf{F}_\text{tx} \mathbf{A}_\text{zf}\mathbf{R}_{\hat{\mathbf{x}}} \mathbf{A}_\text{zf}^{H}\mathbf{F}^H_\text{tx}+\epsilon^2_{\mathbf{t}} \mathbf{I}$. Using the fact that the function $f=tr\{ \mathbf{Z} \mathbf{A}_0 \mathbf{Z}^H \mathbf{A}_1\}$ has derivative given by $\frac{d}{d\mathbf{Z}^{\ast}} f = \mathbf{A}_1 \mathbf{Z} \mathbf{A}_0$ \cite[Table 4]{hjorungnes2007complex}, one may easily find $\frac{\partial }{\partial\mathbf{F}_\text{tx}^{\ast}} tr\{\mathbf{Q}\}=\mathbf{0}$ and $\frac{\partial}{\partial \mathbf{F}_\text{rx}^{\ast}} tr\{\mathbf{Q}\} = \mathbf{0}$. Fixing $\mathbf{F}_\text{tx}$ comes
\begin{equation}
\begin{split}
\frac{\partial}{\partial \mathbf{F}_\text{rx}^{\ast}} tr\{\mathbf{Q}\} = & \mathbf{F}_\text{rx} (p_\text{R} \tilde{\mathbf{H}}_\text{LI}\mathbf{R}_{t} \tilde{\mathbf{H}}_\text{LI}^H+\mathbf{R}_{\mathbf{n}_r})\\
&-(\mathbf{I}-\mathbf{F}_\text{rx})\tilde{\mathbf{G}}_\text{SR}\mathbf{D}_{p_\text{S}} \tilde{\mathbf{G}}_\text{SR}^H = \mathbf{0},
\end{split}
\end{equation}
which, when solved, 
gives \eqref{mmsefilter}. Fixing $\mathbf{F}_\text{rx}$ leads to
\begin{equation}
\begin{split}
\frac{\partial}{\partial \mathbf{F}_\text{tx}^{\ast}} tr\{\mathbf{Q}\} = & \mathbf{D}_{p_\text{S}} \mathbf{F}_\text{rx} \tilde{\mathbf{H}}_\text{LI} (p_\text{R} \mathbf{F}_\text{rx}^H \tilde{\mathbf{H}}_\text{LI}^H\mathbf{F}_\text{tx} \mathbf{A}_{zf} \mathbf{R}_{\hat{\mathbf{x}}}\mathbf{A}_\text{zf}^H)\\
=&\mathbf{D}_{p_\text{S}} \mathbf{F}_\text{rx} \tilde{\mathbf{H}}_\text{LI} (\mathbf{F}_\text{rx} \tilde{\mathbf{H}}_\text{LI} \mathbf{F}_\text{tx})\mathbf{A}_\text{zf} \mathbf{R}_{\hat{\mathbf{x}}}\mathbf{A}_\text{zf}^H = \mathbf{0},
\end{split}
\end{equation}
whose solution is given by \eqref{nsp}.

\section*{Acknowledgments}
\footnotesize This work was funded by FCT (Foundation for Science and Technology) and Instituto de Telecomunicações under project UID/EEA/50008/2013.

%
%
%

\bibliography{biblio}{}
\bibliographystyle{IEEEtran}

\end{document}